\documentclass[journal]{IEEEtran}
\usepackage{ifpdf}

\usepackage{cite}

\ifCLASSINFOpdf
  \usepackage[pdftex]{graphicx}
\else
  \usepackage[dvips]{graphicx}
\fi
\usepackage{amsmath}
\usepackage{array}
\usepackage{fixltx2e}

\usepackage{stfloats}
\usepackage{url}

\usepackage{amssymb} 
\usepackage{mathtools, nccmath}
\usepackage{algorithm}
\usepackage{algpseudocode}
\usepackage{multirow}
\usepackage{color}
\usepackage{mathrsfs}
\usepackage[table, dvipsnames]{xcolor}
\usepackage{color, colortbl}
\definecolor{LightCyan}{rgb}{0.88,1,1}
\usepackage{tikz} 
\usepackage{pgfplots}
\pgfplotsset{compat=1.5}
\usepackage{amsmath,bm}

\begin{document}
	
\markboth{}%
{Shell \MakeLowercase{\textit{et al.}}: Bare Demo of IEEEtran.cls for IEEE Journals}
%
	
%
\title{Q-LIC: Quantizing Learned Image Compression with Channel Splitting}
%
%
%

\author{Heming~Sun,
        Lu~Yu,
        and~Jiro~Katto
\thanks{Heming Sun is with Waseda University, Japan and JST, Japan. (hemingsun@aoni.waseda.jp)}
\thanks{Lu Yu is with Zhejiang University, China. (yul@zju.edu.cn)}
\thanks{Jiro Katto is Waseda University, Japan. (katto@waseda.jp)}
}

%
%



\maketitle

\begin{abstract}
Learned image compression (LIC) has reached a comparable coding gain with traditional hand-crafted methods such as VVC intra. However, the large network complexity prohibits the usage of LIC on resource-limited embedded systems. Network quantization is an efficient way to reduce the network burden. This paper presents a quantized LIC (QLIC) by channel splitting. First, we explore that the influence of quantization error to the reconstruction error is different for various channels. Second, we split the channels whose quantization has larger influence to the reconstruction error. After the splitting, the dynamic range of channels is reduced so that the quantization error can be reduced. Finally, we prune several channels to keep the number of overall channels as origin. By using the proposal, in the case of 8-bit quantization for weight and activation of both main and hyper path, we can reduce the BD-rate by 0.61\%-4.74\% compared with the previous QLIC. Besides, we can reach better coding gain compared with the state-of-the-art network quantization method when quantizing MS-SSIM models. Moreover, our proposal can be combined with other network quantization methods to further improve the coding gain. The moderate coding loss caused by the quantization validates the feasibility of the hardware implementation for QLIC in the future.

\end{abstract}


%
\IEEEpeerreviewmaketitle

\section{Introduction}

\IEEEPARstart{I}{mage} compression is important to relieve the burden of the image transmission and storage. In the past decades, several standards have been developed such as JPEG \cite{Wallace1992TheJS}, JPEG2000 \cite{rabbani2002overview}, WebP \cite{lian2012webp} and HEVC intra (BPG) \cite{sullivan2012overview}. For these standardized methods, the coding components are fixed which are composed of intra prediction, linear transform, quantization and entropy coding. To improve the coding gain, new features such as more intra modes and larger transform kernels have been developed. However, the coding components are optimized separately thus the joint optimization might lead to a higher compression ratio. In addition, the transform kernels such as discrete cosine transform (DCT) are performed linearly, thus adopting the non-linear feature is potential for improving the compression ability.

In the recent years, learned image compression (LIC) has illustrated a superior compression ability. One classical LIC framework is hyper-prior model \cite{balle2018variational} which is composed of main path and hyper path. Main path is served as a non-linear transform, while hyper path is used for the entropy coding. Based on the hyper-prior framework, there have been quite a few of works \cite{minnen2018joint,cheng2020learned,guo2021causal,chen2021end,hu2021learning,wang2020ensemble} improving the coding gain by enhancing the transform ability in the main path and attempting more sophisticated entropy model in the hyper path. As a result, the recent LIC work \cite{guo2021causal} can reach a comparable coding gain with the latest traditional standard VVC intra \cite{bross2021overview}.

Along with a considerable coding gain improvement, the network complexity of LIC also increases hugely as reported in \cite{ma2019image}. First, to store and transfer a huge amount of parameters and activations, the memory consumption is high. Second, the arithmetic operation of weights and activations is computationally intensive when using 32-bit floating point. To alleviate the above two problems, network pruning and quantization are two mainstream approaches. By using the pruning methods in \cite{han2015learning, li2016pruning, he2017channel, luo2017thinet}, some weights and activations are cut so that the memory consumption can be saved. However, the computational cost in the worst case still remains the same as origin which requires the 32-bit floating-point multiplier and adder. By using the quantization methods, the bit-depth of weights and activations is reduced to save the memory consumption. Moreover, the computational burden can also be relieved. As a result, quantized model can be mapped on some specific hardwares \cite{zhang2018dnnbuilder,chen2016eyeriss,zhang2020dnnexplorer,ye2020hybriddnn,jo2017dsip} for the acceleration.

Network quantization can be grouped to two categories according to the quantization targets. First category is to quantize the weight. Han \textit{et al.}\cite{han2015deep} clustered weights and then applied a non-uniform quantization. An incremental network quantization scheme was proposed in \cite{zhou2017incremental} and the key concept was to quantize the partial weights and then compensate the quantization loss by fine tuning the remaining weights recursively, until all the weights have been quantized. Vector quantization was utilized to compress the weights in \cite{gong2014compressing}. Lipschitz constraint-based width-level and multi-level network quantization were proposed for high-bit and low-bit quantization respectively in \cite{xu2019iterative}.

\begin{figure*}[htb]
	\centering
	\includegraphics[height=8.0cm]{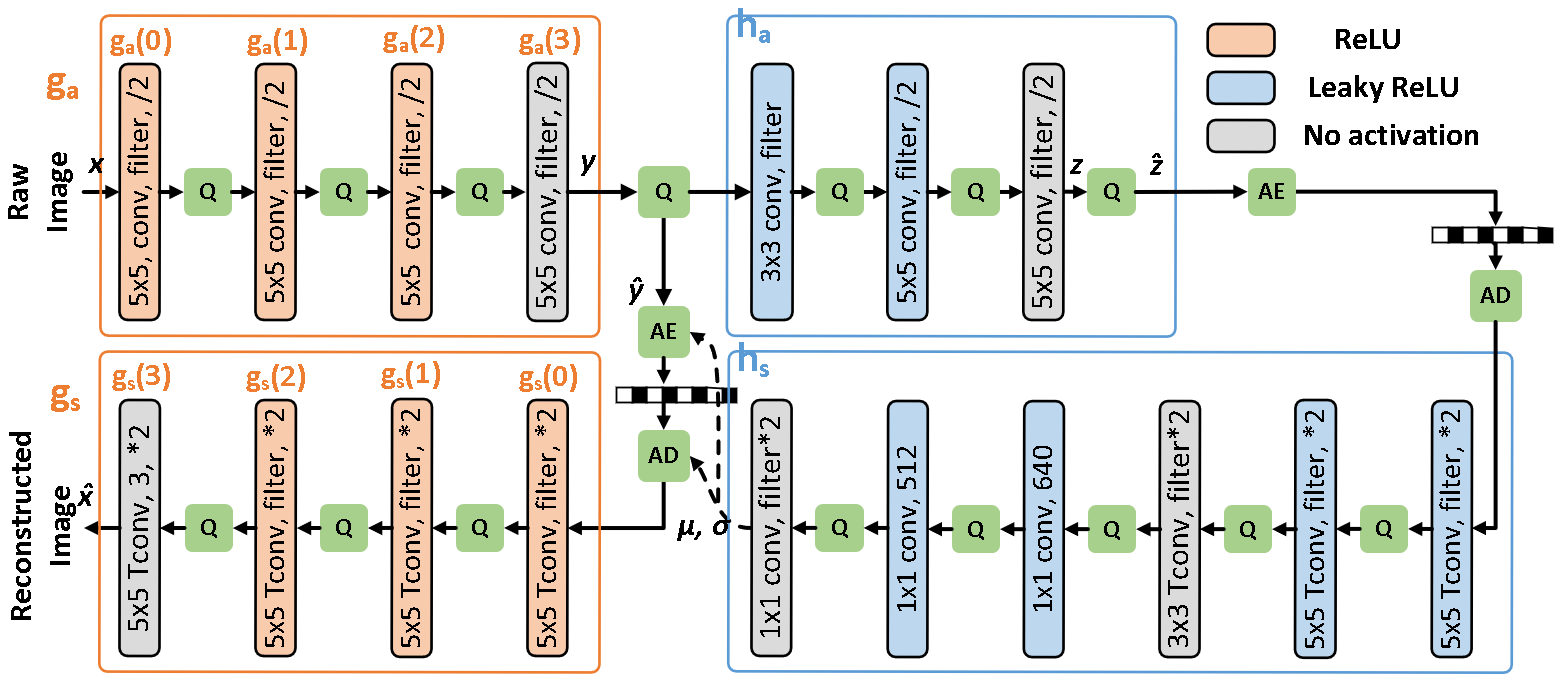}
	\caption{Network model used in this paper. All the weights and activations are quantized. All the arithmetic operations in the network can be performed as fixed-point operations.}
	\label{fig_hyper}
\end{figure*}

Second category is focused on the activation quantization. Park \textit{et al.}\cite{park2018value} presented value-aware quantization so that lower quantization precision (QP) is applied to smaller values. The author also exploited the concept of weighted entropy and analyzed the impact of different weight/activation value on the final accuracy in \cite{park2017weighted}. Mishra \textit{et al.}\cite{mishra2017wrpn} proposed a low precision network providing more filters than the origin which can surpass the accuracy of the baseline full-precision network. There are also several works emphasizing on optimizing the activation range. Jung \textit{et al.} \cite{jung2019learning} determined the efficient activation interval by learning the center and distance parameter. Variants of ReLU with different bounding value and piece-wise function was proposed in \cite{cai2017deep}. Batch normalization \cite{ioffe2015batch} was utilized to purposely generate a Gaussian-like distribution and Lloyd's algorithm \cite{lloyd1982least} was used to optimize the quantization steps. Different with \cite{cai2017deep} whose clipping value is fixed, Choi \textit{et al.}\cite{choi2018pact} optimized the clipping range of activations through training and then linearly quantized both weights and activations to 4-bit. Zhou \textit{et al.} \cite{zhou2019progressive} equipped a low-precision network with a full-precision network and then gradually remove the impact of the full-precision network during the training. \cite{lee2021network} proposed an element-wise gradient scaling to replace the straight-through-estimator in the training.

Though there have been tremendous quantization works, most of them are for the high-level vision tasks such as image recognition and object detection. Up to now, there are very few works for quantizing LIC. \cite{balle2018integer} developed a very heuristic method to train an integer LIC from scratch. \cite{sun2020end} proposed a weight clipping method to reduce the weight quantization error, and its advanced version with a layer-by-layer weight fine tuning was presented in \cite{sun2021learned}. \cite{hong2020efficient} proposed a range pre-processing to bound the dynamic range, and then performed a range-adaptive quantization. For \cite{sun2020end,sun2021learned}, though weight can be quantized with small coding loss, the activation in the main path is not taken into consideration for the quantization. For \cite{balle2018integer,hong2020efficient}, both weights and activations in the main and hyper path are quantized. However, the coding loss is relatively large compared with the non-quantized floating-point baseline.

This paper reduces the coding loss of a quantized LIC (QLIC) framework whose weights and activations are all quantized. The contribution is threefold: 1) For each channel, we formulate the influence of the quantization error to the reconstruction error as $B_k$, and explore that $B_k$ is different for different channels of intermediate layers. 2) We split the channels with large $B_k$ to reduce the dynamic range, so that the quantization loss can be reduced with the same quantization bit budget. 3) To keep the overall number of channels as origin, the same number of split channels are pruned. Energy is used as the metric for selecting the pruned channels. The experiments show that when quantizing all the weights and activations to 8-bit, the BD-rate can be saved by 4.74\% for MS-SSIM model compared with the previous work. In addition, compared with the state-of-the-art network quantization work \cite{lee2021network}, we can also achieve better results. Furthermore, we also illustrate that the combination of our method and \cite{lee2021network} can further improve the coding gain.

\section{Quantized Learned Image Compression}

\subsection{Quantization Method}

Quantization can be classified into linear quantization (LQ) and non-linear quantization (NLQ). LQ is shown in the below
\begin{equation}
	\small
	Q(x;s) = \begin{cases}
		\lfloor \tfrac{x}{s} \rceil \cdot s, & \text{if } n \leq \lfloor \tfrac{x}{s} \rceil \leq p \\
		n \cdot s, & \text{if } \lfloor \tfrac{x}{s} \rceil < n\\
		p \cdot s, & \text{if } \lfloor \tfrac{x}{s} \rceil > p\\
	\end{cases}
	\label{eq_lq}
\end{equation}
\noindent where $s$ is the quantization precision which is related with the dynamic range $t$, $p$ and $n$ are the positive and negative bound integers which are based on the bit budget $b$ and the type of activation functions. For unsigned activation functions such as ReLU, $s=\tfrac{2^{\lceil log2^t \rceil}}{2^b}$, $p=2^b-1$ and $n=0$. In the case of signed activation functions such as Leaky-ReLU, $s=\tfrac{2^{\lceil log2^t \rceil}}{2^{b-1}}$, $p=2^{b-1}-1$ and $n=-2^{b-1}$.

Different from LQ, the quantization precision is non-uniform for NLQ. Optimal quantization precision will be decided according to the value distribution. For instance, in the case of bell-shape distribution, higher precision will be used for the smaller values.

As described in the above, the dynamic range $t$ will decide the quantization precision $s$. However, the dynamic range of different activation outputs are different. Therefore, we need to decide the grouping scheme for $t$. To reduce the quantization error, each value has its own $t$. To reduce the storage burden, all the values can share one $t$. Considering the structural feature of convolution neural network, appropriate grouping scheme for $t$ can be performed in the channel-wise or layer-wise.

\subsection{Quantized LIC}

In this work, we build the baseline network based on the \textit{hyperprior-5} model in \cite{cheng2019deep} as shown in Fig. \ref{fig_hyper}. There are four layers in the main analysis transform $g_a$ and synthesis transform $g_s$ respectively as described in the following equation
\begin{equation}
\bm{y}=g_a(\bm{x})=g^{(3)}_a \circ g^{(2)}_a \circ g^{(1)}_a \circ g^{(0)}_a (\bm{x})
\label{eq_at}
\end{equation}
\begin{equation}
\hat{\bm{x}}=g_s(\hat{\bm{y}})=g^{(3)}_s \circ g^{(2)}_s \circ g^{(1)}_s \circ g^{(0)}_s (\hat{\bm{y}})
\label{eq_st}
\end{equation}
\noindent where $\hat{\bm{y}}=\lfloor \bm{y} \rceil$. The hyper path can be formulated as below
\begin{equation}
	\bm{z}=h_a(\hat{\bm{y}})
	\label{eq_ha}
\end{equation}
\begin{equation}
	\bm{\mu_y},\bm{\sigma_y}=h_s(\hat{\bm{z}})
	\label{eq_hs}
\end{equation}
\noindent where $\hat{\bm{z}}=\lfloor \bm{z} \rceil$. For the activation function, ReLU and Leaky-ReLU rather than the generalized divisive normalization (GDN) \cite{balle2016end} is used to ease the hardware implementation in the future. The slope of Leaky-ReLU is set as 0.125 which can be operated as right shifting.

The target of this paper is to quantize the weights and activations of all the layers. About the weight quantization, previous work \cite{sun2021learned} has illustrated that linearly quantizing to 8-bit will not lead to large coding loss. About the activation quantization, quantizing the hyper path will not cause coding loss as reported in \cite{sun2021learned,balle2018integer}. Therefore, the remaining issue is to quantize the activation in the main path, including both bottleneck layer and non-bottleneck layers.

For the outputs of bottleneck layer $g^{(3)}_a$, without any bit-budget limitation, it is just rounded thus the quantization precision is one. In the case of limited bit-budget, in order not to clip the dynamic range, the quantization precision can be decided by the following equation.
\begin{equation}
\small
s = \begin{cases}
1 , & \text{if } 2^{\lceil log2^{t} \rceil} < 2^{b-1} \\
\dfrac{2^{\lceil log2^t \rceil}}{2^{b-1}} , & \text{if } 2^{\lceil log2^{t} \rceil} \geq 2^{b-1} \\
\end{cases}
\label{eq_at3}
\end{equation}

\begin{figure*}[htb]
	\centering
	\includegraphics[height=3.3cm]{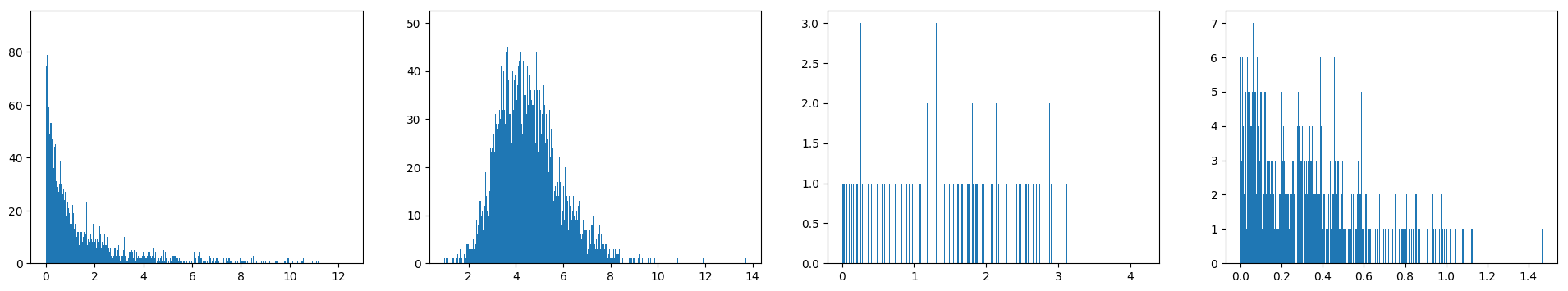}
	\caption{Activation distribution for four channels of $g^{(0)}_a$. There is no unified distribution for various channels, and dynamic range also varies for different channels.}
	\label{fig_at0}
\end{figure*}

For the outputs of non-bottleneck layers, one distribution instance for $g^{(0)}_a$ is shown in Fig. \ref{fig_at0}. We can see that there is no obvious distribution shape, thus we use LQ rather than NLQ. Besides, the dynamic range of various channels are different, so channel-wise rather than layer-wise quantization is adopted.

\section{Relationship Between Quantization Error and R-D Cost}

\subsection{Reconstruction Error for Linear System}

According to \cite{katto1991performance}, for the linear sub-band coding system, when quantizing the outputs of analysis transform, we have the following relationship
\begin{equation}
\sigma_r^2=\sum_{k=0}^{K-1}B_k\sigma_{q_k}^2
\label{eq_var_r}
\end{equation}
\noindent where $\sigma_{q_k}^2$ is the variance of the quantization error and $\sigma_r^2$ is the variance of the reconstruction error, $K$ is the number of channels. By substituting the approximate relationship \cite{jayant1984digital}, 
\begin{equation}
\sigma_{q_k}^2\simeq\epsilon_k^2 2^{-2b_k}\sigma_{y_k}^2
\label{eq_var_qk}
\end{equation}
\noindent where $y_k$ is the outputs of analysis transform, $\epsilon_k$ is related with the distribution of $y_k$ and $\sigma_{y_k}$ is the variance of $y_k$, $b_k$ is the quantization bit for each channel.

\subsection{Reconstruction Error for Quantized LIC}

LIC has two differences from linear sub-band coding system. First difference is that LIC is a non-linear system when using non-linear activation functions. Second difference is that in \cite{katto1991performance}, only the outputs of the bottleneck layer are quantized, while outputs of both bottleneck layer and non-bottleneck layer are quantized in our quantized LIC.

For the $g$-th non-bottleneck layer, the relationship between the variance of the quantization error and the variance of the reconstruction error can be written as below.
\begin{equation}
\sigma_{r(g)}^2=\sum_{k=0}^{K-1}B^{(g)}_k\sigma_{q^{(g)}_k}^2
\label{eq_var_r_lic}
\end{equation}

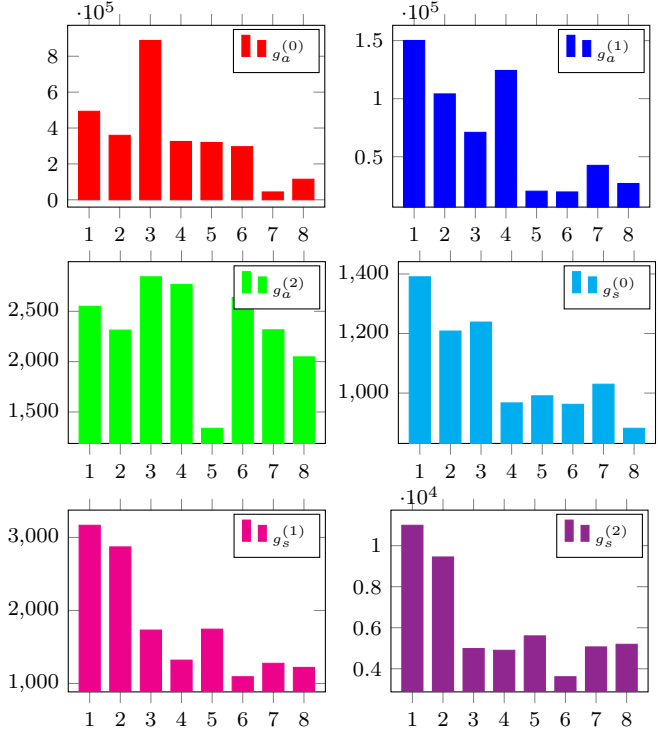
\begin{figure}[tb]
	\centering
	\leftskip-0.0cm
	\hspace{4mm}
	\begin{minipage}[b]{0.4\linewidth}
		\begin{tikzpicture}
			\begin{axis}[ybar, bar width=8pt, legend columns=4,legend style={font=\tiny},font=\footnotesize,width=5.0cm,height=4cm,xtick=data,legend columns=6,yticklabel style={/pgf/number format/fixed}]
				\addplot [red,fill=red] table {at0_bk.dat};
				\legend{$g^{(0)}_a$}
			\end{axis}
		\end{tikzpicture}
	\end{minipage}
	\hspace{3mm}
	\begin{minipage}[b]{0.4\linewidth}
		\begin{tikzpicture}
			\begin{axis}[ybar, bar width=8pt, legend columns=4,legend style={font=\tiny},font=\footnotesize,width=5.0cm,height=4cm,xtick=data,legend columns=6,yticklabel style={/pgf/number format/precision=3}]
				\addplot [blue,fill=blue] table {at1_bk.dat};
				\legend{$g^{(1)}_a$}
			\end{axis}
		\end{tikzpicture}
		
	\end{minipage}
	\begin{minipage}[b]{0.4\linewidth}
		\begin{tikzpicture}
			\begin{axis}[ybar, bar width=8pt, legend columns=4,legend style={font=\tiny},font=\footnotesize,width=5.0cm,height=4cm,xtick=data,legend columns=6,yticklabel style={    /pgf/number format/.cd,
					fixed,
					precision=1}]
			\addplot [green,fill=green] table {at2_bk.dat};
			\legend{$g^{(2)}_a$}
			\end{axis}
		\end{tikzpicture}
		
	\end{minipage}
	\hspace{6mm}
	\begin{minipage}[b]{0.4\linewidth}
		\begin{tikzpicture}
			\begin{axis}[ybar, bar width=8pt, legend columns=4,legend style={font=\tiny},font=\footnotesize,width=5.0cm,height=4cm,xtick=data,legend columns=6,yticklabel style={/pgf/number format/precision=3}]
				\addplot [cyan,fill=cyan] table {st0_bk.dat};
				\legend{$g^{(0)}_s$}
			\end{axis}
		\end{tikzpicture}
		
	\end{minipage}

	\begin{minipage}[b]{0.4\linewidth}	
		\begin{tikzpicture}
			\begin{axis}[ybar, bar width=8pt, legend columns=4,legend style={font=\tiny},font=\footnotesize,width=5.0cm,height=4cm,xtick=data,legend columns=6,yticklabel style={/pgf/number format/.cd,fixed,precision=1}]
				\addplot [Magenta,fill=Magenta] table {st1_bk.dat};
				\legend{$g^{(1)}_s$}
			\end{axis}
		\end{tikzpicture}
	\end{minipage}
	\hspace{8mm}
	\begin{minipage}[b]{0.4\linewidth}
		\begin{tikzpicture}
			\begin{axis}[ybar, bar width=8pt, legend columns=4,legend style={font=\tiny},font=\footnotesize,width=5.0cm,height=4cm,xtick=data,legend columns=6,yticklabel style={/pgf/number format/precision=3}]
				\addplot [Plum,fill=Plum] table {st2_bk.dat};
				\legend{$g^{(2)}_s$}
			\end{axis}
		\end{tikzpicture}
	\end{minipage}
	
	\caption{$B_k$ for eight channels with the largest energy of different non-bottleneck layers.}
	\label{fig_bk}
\end{figure}

To evaluate $B_k$, in addition to the bottleneck layer, we also quantize the $k$-th channel of the $g$-th layer. In this case, the reconstruction error can be written as
\begin{equation}
\sigma_{r}^2=B^{(g)}_k\sigma_{q^{(g)}_k}^2 + \sigma_{r ( g_a^{(3)} ) }^2
\label{eq_kk}
\end{equation}
\noindent where $\sigma_{r ( g_a^{(3)} ) }^2$ is the reconstruction error caused by quantizing the bottleneck layer. We quantize the $k$-th channel with five different quantization bits, and obtain $B_k$ by the linear regression. Results are shown in Fig. \ref{fig_bk}. We can see that $B_k$ for various channels are different. Large $B_k$ means that the quantization error will have more influence to the reconstruction error.

Now that when quantizing each individual layer, the quantization influence to the reconstruction error of each channel can be evaluated by $B_k$, then we explore the quantization influence of each layer to the reconstruction error when quantizing multiple layers. We first quantize the $g$-th non-bottleneck layer as well as the bottleneck layer to obtain $\sigma_{r(g)}^2$ as below.
\begin{equation}
\sigma_{r(g)}^2=\sigma_{r}^2 - \sigma_{r ( g_a^{(3)} ) }^2
\label{eq_vv}
\end{equation}

\begin{figure}[h]
	\footnotesize
	\centering
	\leftskip-0.0cm
	
	\pgfplotsset{
		compat=1.11,
		legend image code/.code={
			\draw[mark repeat=2,mark phase=2]
			plot coordinates {
				(0cm,0cm)
				(0.15cm,0cm)        
				(0.3cm,0cm)         
			};%
		}
	}
	
	\begin{minipage}[b]{0.4\linewidth}
		\pgfplotsset{
		}
		\pgfplotsset{every axis plot/.append style={line width=0.7pt}}
		\tikzset{every mark/.append style={scale=0.5}}
		\begin{tikzpicture}
			\begin{axis}[
				width=4.5cm, height=4.5cm,
				x label style={at={(axis description cs:0.5,-0.1)},anchor=north},
				y label style={at={(axis description cs:-0.05,.5)},anchor=south},
				y tick label style = { /pgf/number format/.cd, precision=3, /tikz/.cd},
				minor y tick num = 2,
				grid = both,
				grid style = {gray!30},
				legend entries = {y=1.034x+69.35},
				legend style={font=\fontsize{6}{6}\selectfont, row sep=-3.5pt, at={(0.5,-0.26)}, anchor=north, inner xsep=0pt,inner ysep=0pt},
				legend cell align=left,
				legend pos = {north west},
				]
				\addplot[smooth, red, mark=square*, mark size=2.5pt] coordinates {
					(0.6516718999999966,   70.03493585)
					(2.5296005400000467,   71.84689607)
					(9.988277400000015 ,  79.81281641)
					(41.821252800000025 ,  112.5783365)	
				};
			\end{axis}
		\end{tikzpicture}
		
		
	\end{minipage}\hspace{4mm}
	\begin{minipage}[b]{0.4\linewidth}
		\pgfplotsset{
		}
		\pgfplotsset{every axis plot/.append style={line width=0.7pt}}
		\tikzset{every mark/.append style={scale=0.5}}
		\begin{tikzpicture}
			\begin{axis}[
				width=4.5cm, height=4.5cm,
				x label style={at={(axis description cs:0.5,-0.1)},anchor=north},
				y label style={at={(axis description cs:-0.05,.5)},anchor=south},
				y tick label style = { /pgf/number format/.cd, precision=3, /tikz/.cd},
				minor y tick num = 2,
				grid = both,
				grid style = {gray!30},
				legend entries = {y=1.066x+54.07},
				legend style={font=\fontsize{6}{6}\selectfont, row sep=-3.5pt, at={(0.5,-0.26)}, anchor=north, inner xsep=0pt,inner ysep=0pt},
				legend cell align=left,
				legend pos = {north west},
				]
				\addplot[smooth, blue, mark=square*, mark size=2.5pt] coordinates {
				(0.776936830000011 ,  55.14113066)
				(3.3012714500000016 ,  57.72953182)
				(13.629772830000036  , 68.09355524)
				(56.34627914999999  , 114.21875339)
				};
			\end{axis}
		\end{tikzpicture}
		
		
	\end{minipage}\hspace{4mm}
	\begin{minipage}[b]{0.4\linewidth}
		\pgfplotsset{
		}
		\pgfplotsset{every axis plot/.append style={line width=0.7pt}}
		\tikzset{every mark/.append style={scale=0.5}}
		\begin{tikzpicture}
			\begin{axis}[
				width=4.5cm, height=4.5cm,
				x label style={at={(axis description cs:0.5,-0.1)},anchor=north},
				y label style={at={(axis description cs:-0.05,.5)},anchor=south},
				y tick label style = { /pgf/number format/.cd, precision=3, /tikz/.cd},
				minor y tick num = 2,
				grid = both,
				grid style = {gray!30},
				legend entries = {y=1.045x+48.68},
				legend style={font=\fontsize{6}{6}\selectfont, row sep=-3.5pt, at={(0.5,-0.26)}, anchor=north, inner xsep=0pt,inner ysep=0pt},
				legend cell align=left,
				legend pos = {north west},
				]
				\addplot[smooth, green, mark=square*, mark size=2.5pt] coordinates {
				(0.9164611200000081,   49.70419047)
				(3.327101500000026 ,  52.21906217)
				(13.859368670000038 ,  62.98317273)
				(57.088679660000025 ,  108.35377375)
				};
			\end{axis}
		\end{tikzpicture}
		
		
	\end{minipage}\hspace{4mm}
	\begin{minipage}[b]{0.4\linewidth}
	\pgfplotsset{
	}
	\pgfplotsset{every axis plot/.append style={line width=0.7pt}}
	\tikzset{every mark/.append style={scale=0.5}}
	\begin{tikzpicture}
		\begin{axis}[
				width=4.5cm, height=4.5cm,
			x label style={at={(axis description cs:0.5,-0.1)},anchor=north},
			y label style={at={(axis description cs:-0.05,.5)},anchor=south},
			y tick label style = { /pgf/number format/.cd, precision=3, /tikz/.cd},
			minor y tick num = 2,
			grid = both,
			grid style = {gray!30},
			legend entries = {y=1.049x+202.16},
			legend style={font=\fontsize{6}{6}\selectfont, row sep=-3.5pt, at={(0.5,-0.26)}, anchor=north, inner xsep=0pt,inner ysep=0pt},
			legend cell align=left,
			legend pos = {north west},
			]
			\addplot[smooth, cyan, mark=square*, mark size=2.5pt] coordinates {
			(0.7031071800001882,   202.65972095)
			(1.534467480000103 ,  203.89006424)
			(6.670207960000198 ,  209.3265241)
			(29.03960671000027 ,  232.60006671)
			};
		\end{axis}
	\end{tikzpicture}
	
	
	\end{minipage}\hspace{4mm}
	\begin{minipage}[b]{0.4\linewidth}
	\pgfplotsset{
	}
	\pgfplotsset{every axis plot/.append style={line width=0.7pt}}
	\tikzset{every mark/.append style={scale=0.5}}
	\begin{tikzpicture}
		\begin{axis}[
			width=4.5cm, height=4.5cm,
			x label style={at={(axis description cs:0.5,-0.1)},anchor=north},
			y label style={at={(axis description cs:-0.05,.5)},anchor=south},
			y tick label style = { /pgf/number format/.cd, precision=3, /tikz/.cd},
			minor y tick num = 2,
			grid = both,
			grid style = {gray!30},
			legend entries = {y=1.084x+139.26},
			legend style={font=\fontsize{6}{6}\selectfont, row sep=-3.5pt, at={(0.5,-0.26)}, anchor=north, inner xsep=0pt,inner ysep=0pt},
			legend cell align=left,
			legend pos = {north west},
			]
			\addplot[smooth, Magenta, mark=square*, mark size=2.5pt] coordinates {
			(0.5344961299999795,   140.22526614)
			(2.475054030000024 ,  142.01873885)
			(9.956239509999932  , 149.44763056)
			(41.328467509999996 ,  184.19448302)
			};
		\end{axis}
	\end{tikzpicture}
	
	
	\end{minipage}\hspace{4mm}
	\begin{minipage}[b]{0.4\linewidth}
	\pgfplotsset{
	}
	\pgfplotsset{every axis plot/.append style={line width=0.7pt}}
	\tikzset{every mark/.append style={scale=0.5}}
	\begin{tikzpicture}
		\begin{axis}[
			width=4.5cm, height=4.5cm,
			x label style={at={(axis description cs:0.5,-0.1)},anchor=north},
			y label style={at={(axis description cs:-0.05,.5)},anchor=south},
			y tick label style = { /pgf/number format/.cd, precision=3, /tikz/.cd},
			minor y tick num = 2,
			grid = both,
			grid style = {gray!30},
			legend entries = {y=1.038x+118.},
			legend style={font=\fontsize{6}{6}\selectfont, row sep=-3.5pt, at={(0.5,-0.26)}, anchor=north, inner xsep=0pt,inner ysep=0pt},
			legend cell align=left,
			legend pos = {north west},
			]
			\addplot[smooth, Plum, mark=square*, mark size=2.5pt] coordinates {
			(0.9499018100000285,   119.6736126)
			(2.8060525299998744 ,  121.65885311)
			(10.640213249999988 ,  129.58963924)
			(44.120103210000025 ,  164.48097526)
			};
		\end{axis}
	\end{tikzpicture}
	
	
	\end{minipage}

	\caption{The relationship between $\sum^{g\in G} \sigma_{r(g)}^2$ and $\sigma_{r}^2$. $G$ includes all the non-bottleneck layers.}
	\label{fig_mse_sum}
	
\end{figure}
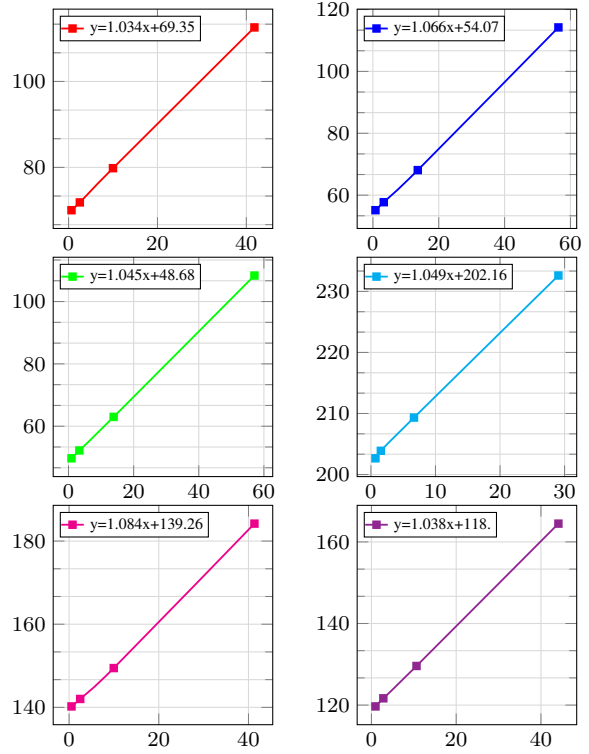

After that, we quantize all the layers to obtain reconstruction error $\sigma_{r}^2$. The relationship between $\sum^{g\in G} \sigma_{r(g)}^2$ and $\sigma_{r}^2$ is shown in Fig. \ref{fig_mse_sum}. From the results, we can clarify that the reconstruction error caused by quantizing each layer can be simply accumulated as below
\begin{equation}
\begin{aligned}
\sigma_{r}^2&=\sum^{g\in G} \sigma_{r(g)}^2 + \sigma_{r ( g_a^{(3)} ) }^2
\end{aligned}
\label{eq_ck}
\end{equation}
\noindent where $G$ is all the non-bottleneck layers including $g_a^{(0,1,2)}$ and $g_s^{(0,1,2)}$. Therefore, to reduce the overall reconstruction error, we should reduce the quantization error for each layer.

\subsection{Rate-Distortion for Quantized LIC}

The relationship between the quantization error and reconstruction error has been given in the above. We analyze Rate-distortion (R-D) cost in this section. R-D cost is calculated by 
\begin{equation}
J=R+\lambda D
\end{equation}
\noindent where $R$ is the rate and $D$ is the distortion. In the case of distortion being MSE, given that the reconstruction error is a zero-mean uniform distribution, $\sigma_{r}^2$ can be directly used as the distortion. Therefore, R-D cost can be rewritten as
\begin{equation}
\begin{aligned}
J&=R+\lambda \sigma_{r}^2 \\
&=R+\lambda (\sum^{g\in G}\sigma_{r(g)}^2 + \sigma_{r ( g_a^{(3)} ) }^2  )\\
&=J_0 + \lambda \sum^{g\in G}\sigma_{r(g)}^2 \\
&=J_0 + \lambda \sum^{g\in G}\sum_{k=0}^{K-1}B^{(g)}_k\sigma_{q^{(g)}_k}^2
\end{aligned}
\label{eq_mse}
\end{equation}

\noindent where the former part $J_0$ is the R-D cost in the case of only quantizing the bottleneck layer and the latter part represents the R-D cost increase when quantizing the non-bottleneck layers.

\begin{figure}[h]
	\footnotesize
	\centering
	\leftskip-0.0cm
	
	\pgfplotsset{
	compat=1.11,
	legend image code/.code={
		\draw[mark repeat=2,mark phase=2]
		plot coordinates {
			(0cm,0cm)
			(0.15cm,0cm)        
			(0.3cm,0cm)         
		};%
	}
}

\begin{minipage}[b]{0.4\linewidth}
	\pgfplotsset{
	}
	\pgfplotsset{every axis plot/.append style={line width=0.7pt}}
	\tikzset{every mark/.append style={scale=0.5}}
	\begin{tikzpicture}
		\begin{axis}[
			width=4.5cm, height=4.5cm,
			x label style={at={(axis description cs:0.5,-0.1)},anchor=north},
			y label style={at={(axis description cs:-0.05,.5)},anchor=south},
			y tick label style = { /pgf/number format/.cd, precision=4, /tikz/.cd},
			minor y tick num = 2,
			grid = both,
			grid style = {gray!30},
			legend entries = {y=-0.0005x+1.01},
			legend style={font=\fontsize{6}{6}\selectfont, row sep=-3.5pt, at={(0.5,-0.26)}, anchor=north, inner xsep=0pt,inner ysep=0pt},
			legend cell align=left,
			legend pos = {south west},
			]
			\addplot[smooth, red, mark=square*, mark size=2.5pt] coordinates {
			(69.39794816 ,  0.97953633)
			(69.57169999 ,  0.97946715)
			(70.13994005 ,  0.97916719)
			};
		\end{axis}
	\end{tikzpicture}
	
	
\end{minipage}\hspace{4mm}
\begin{minipage}[b]{0.4\linewidth}
	\pgfplotsset{
	}
	\pgfplotsset{every axis plot/.append style={line width=0.7pt}}
	\tikzset{every mark/.append style={scale=0.5}}
	\begin{tikzpicture}
		\begin{axis}[
			width=4.5cm, height=4.5cm,
			x label style={at={(axis description cs:0.5,-0.1)},anchor=north},
			y label style={at={(axis description cs:-0.05,.5)},anchor=south},
			y tick label style = { /pgf/number format/.cd, precision=4, /tikz/.cd},
			minor y tick num = 2,
			grid = both,
			grid style = {gray!30},
			legend entries = {y=-0.0005x+1.01},
			legend style={font=\fontsize{6}{6}\selectfont, row sep=-3.5pt, at={(0.5,-0.26)}, anchor=north, inner xsep=0pt,inner ysep=0pt},
			legend cell align=left,
			legend pos = {south west},
			]
			\addplot[smooth, blue, mark=square*, mark size=2.5pt] coordinates {
			(69.47108205 ,  0.97949491)
			(69.75118934 ,  0.97937831)
			(70.78584713 , 0.97883718)
			};
		\end{axis}
	\end{tikzpicture}
	
	
\end{minipage}\hspace{4mm}
\begin{minipage}[b]{0.4\linewidth}
	\pgfplotsset{
	}
	\pgfplotsset{every axis plot/.append style={line width=0.7pt}}
	\tikzset{every mark/.append style={scale=0.5}}
	\begin{tikzpicture}
		\begin{axis}[
			width=4.5cm, height=4.5cm,
			x label style={at={(axis description cs:0.5,-0.1)},anchor=north},
			y label style={at={(axis description cs:-0.05,.5)},anchor=south},
			y tick label style = { /pgf/number format/.cd, precision=4, /tikz/.cd},
			minor y tick num = 2,
			grid = both,
			grid style = {gray!30},
			legend entries = {y=-0.0006x+1.02},
			legend style={font=\fontsize{6}{6}\selectfont, row sep=-3.5pt, at={(0.5,-0.26)}, anchor=north, inner xsep=0pt,inner ysep=0pt},
			legend cell align=left,
			legend pos = {south west},
			]
			\addplot[smooth, green, mark=square*, mark size=2.5pt] coordinates {
			(69.47286436 ,  0.97950927)
			(69.71813181 , 0.97937601)
			(70.75899145 ,  0.97872942)
			};
		\end{axis}
	\end{tikzpicture}
	
	
\end{minipage}\hspace{4mm}
\begin{minipage}[b]{0.4\linewidth}
	\pgfplotsset{
	}
	\pgfplotsset{every axis plot/.append style={line width=0.7pt}}
	\tikzset{every mark/.append style={scale=0.5}}
	\begin{tikzpicture}
		\begin{axis}[
			width=4.5cm, height=4.5cm,
			label style={font=\tiny},
			x label style={at={(axis description cs:0.5,-0.1)},anchor=north},
			y label style={at={(axis description cs:-0.05,.5)},anchor=south},
			y tick label style = { /pgf/number format/.cd, precision=4, /tikz/.cd},
			minor y tick num = 2,
			grid = both,
			grid style = {gray!30},
			legend entries = {y=-0.0007x+1.03},
			legend style={font=\fontsize{6}{6}\selectfont, row sep=-3.5pt, at={(0.5,-0.26)}, anchor=north, inner xsep=0pt,inner ysep=0pt},
			legend cell align=left,
			legend pos = {south west},
			]
			\addplot[smooth, cyan, mark=square*, mark size=2.5pt] coordinates {
			(69.551221  , 0.97941586)
			(70.06969558 ,  0.97903671)
			(72.06002659 ,  0.97764287)
			};
		\end{axis}
	\end{tikzpicture}
	
	
\end{minipage}\hspace{4mm}
\begin{minipage}[b]{0.4\linewidth}
	\pgfplotsset{
	}
	\pgfplotsset{every axis plot/.append style={line width=0.7pt}}
	\tikzset{every mark/.append style={scale=0.5}}
	\begin{tikzpicture}
		\begin{axis}[
			width=4.5cm, height=4.5cm,
			x label style={at={(axis description cs:0.5,-0.1)},anchor=north},
			y label style={at={(axis description cs:-0.05,.5)},anchor=south},
			y tick label style = { /pgf/number format/.cd, precision=4, /tikz/.cd},
			minor y tick num = 2,
			grid = both,
			grid style = {gray!30},
			legend entries = {y=-0.0008x+1.03},
			legend style={font=\fontsize{6}{6}\selectfont, row sep=-3.5pt, at={(0.5,-0.26)}, anchor=north, inner xsep=0pt,inner ysep=0pt},
			legend cell align=left,
			legend pos = {south west},
			]
			\addplot[smooth, Magenta, mark=square*, mark size=2.5pt] coordinates {
			(69.51735454 ,  0.97942707)
			(69.89502398 ,  0.97914226)
			(71.42732196 ,  0.9779142)
			};
		\end{axis}
	\end{tikzpicture}
	
	
\end{minipage}\hspace{4mm}
\begin{minipage}[b]{0.4\linewidth}
	\pgfplotsset{
	}
	\pgfplotsset{every axis plot/.append style={line width=0.7pt}}
	\tikzset{every mark/.append style={scale=0.5}}
	\begin{tikzpicture}
		\begin{axis}[
			width=4.5cm, height=4.5cm,
			x label style={at={(axis description cs:0.5,-0.1)},anchor=north},
			y label style={at={(axis description cs:-0.05,.5)},anchor=south},
			y tick label style = { /pgf/number format/.cd, precision=4, /tikz/.cd},
			minor y tick num = 2,
			grid = both,
			grid style = {gray!30},
			legend entries = {y=-0.0009x+1.04},
			legend style={font=\fontsize{6}{6}\selectfont, row sep=-3.5pt, at={(0.5,-0.26)}, anchor=north, inner xsep=0pt,inner ysep=0pt},
			legend cell align=left,
			legend pos = {south west},
			]
			\addplot[smooth, Plum, mark=square*, mark size=2.5pt] coordinates {
			(69.50608169 ,  0.97943385)
			(69.78906864 , 0.97910329)
			(71.08227115 ,  0.97796346)

			};
		\end{axis}
	\end{tikzpicture}
	
	
\end{minipage}
	
	\caption{The relationship between $\sigma_r^2$ and MS-SSIM for different layers.}
	\label{fig_mse_msssim}
	
\end{figure}
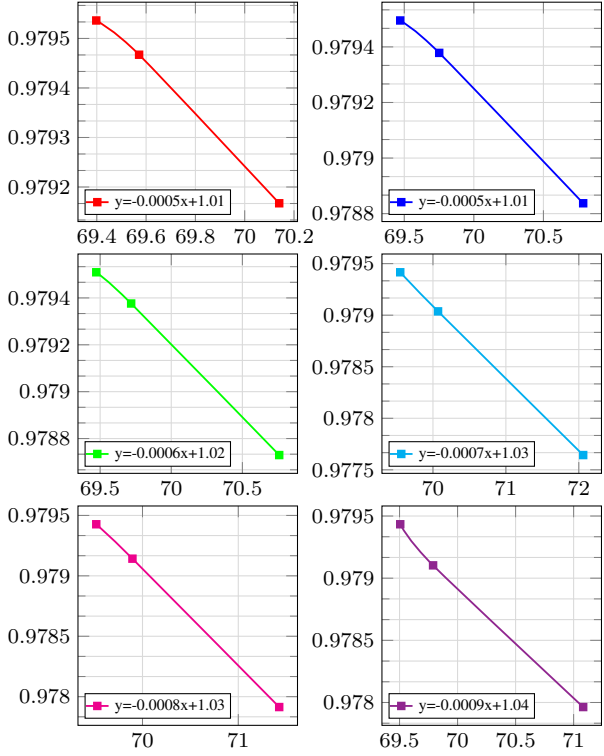

In the case of distortion being $1-\text{MS-SSIM}$, the relationship between $\sigma_{r}^2$ and MS-SSIM in the case of quantizing different layers are shown in Fig. \ref{fig_mse_msssim}. We can see that the relationship can be approximated as below.

\begin{equation}
\text{MS-SSIM} \simeq -a \sigma_{r}^2 + 1
\end{equation}
\noindent where $a$ is a positive value. Therefore, R-D cost can be rewritten as
\begin{equation}
\begin{aligned}
J& \simeq R+ \lambda a \sigma_{r}^2 \\
&=J_0 + \lambda a \sum^{g\in G}\sum_{k=0}^{K-1}B^{(g)}_k\sigma_{q^{(g)}_k}^2
\end{aligned}
\label{eq_msssim}
\end{equation}

From Eq. \ref{eq_mse} and Eq. \ref{eq_msssim}, we can see that when quantizing multiple layers in addition to the bottleneck layer, R-D cost increase is proportional to the quantization error.

Regarding the rate, it will not be changed if we quantize the synthesis transform. However, rate will be different from the non-quantized baseline if we quantize the analysis transform. According to our experimental results, when using 8-bit to quantize the analysis transform, there is no big change for the rate. In addition, rate can be reduced by fine tuning the hyper path. Therefore, we assume that the quantization does not influence the rate in the above derivation.

\section{Proposed Channel Splitting}

As described in the previous section, reducing the quantization error for each layer can contribute to a better coding gain. By splitting the channel, the dynamic range can be shrunk so that the quantization error can be reduced.

In this section, we formulate the channel splitting problem at first, and then propose the channel splitting methods for non-bottleneck and bottleneck layer respectively. In order not to increase the network burden, some channels are pruned. Finally, we present the overall framework to illustrate the procedures of channel splitting and pruning.

\subsection{Channel Splitting Formulation}

For the $i$-th layer, the activation output is $\bm{O}_i \in \mathbb{R}^{w_i \times h_i \times c_i}$ where $w_i$ and $h_i$ are the width and height, and $c_i$ is the number of the output channels. For the $n$-th channel $\bm{O}^{(n)}_{i} \in \mathbb{R}^{w_i \times h_i}$, it is calculated by
\begin{equation}
\small
\bm{O}^{(n)}_{i} = \sum_{m=0}^{C_{i-1}}\bm{O}^{(m)}_{i-1}*\bm{W}^{(m,n)}_{i} + \bm{b}^{(n)}_{i}
\label{eq_1}
\end{equation}
\noindent where $m$ and $n$ are the index of input and output channel, $\bm{O}_{i-1}$ is the output of the previous layer, $\bm{W}^{(m,n)}_{i}$ and $\bm{b}^{(n)}_{i}$ are weight and bias.

Given that a specific channel $\bm{O}^{(n)}_i$ is split to $\mathcal{K}$ channels $\bm{O}^{\prime(n_k)}_{i}$, each split channel can be calculated by the following equation.
\begin{equation}
\small
\begin{aligned}
\bm{O}^{\prime(n_k)}_{i} = \sum_{m=0}^{C_{i-1}}\bm{O}^{(m)}_{i-1}*\bm{W}^{\prime(m,n_k)}_{i} + \bm{b}^{\prime(n_k)}_i
\end{aligned}
\label{eq_2}
\end{equation}

To ensure that the network function does not change after the splitting, we should have the following relationship.
\begin{equation}
\small
\sum_{k=0}^{\mathcal{K}-1}\bm{O}^{\prime(n_k)}_{i} = \bm{O}^{(n)}_{i}
\label{eq_o}
\end{equation}

To satisfy Eq. \ref{eq_o}, the weight and bias for the split channels are set as below
\begin{equation}
\small
\bm{W}^{\prime(:,n_k)}_{i} = \dfrac{\bm{W}^{(:,n)}_{i}}{\mathcal{K}}
\label{eq_w}
\end{equation}
\begin{equation}
\small
\bm{b}^{\prime(n_k)}_{i} = \dfrac{\bm{b}^{(n)}_{i}}{\mathcal{K}} + \delta_k
\label{eq_b}
\end{equation}

\noindent where $:$ is a wild card for all the input channels, and $\delta_k$ is a small offset to generate various quantized value for split channels. As long as $\sum_{k=0}^{\mathcal{K}-1}\delta_k=0$, we can satisfy Eq. \ref{eq_o} based on Eq. \ref{eq_w} and Eq. \ref{eq_b}.

The split output channels of the $i$-th layer are also the input channels of the $i$+1-th layer. Therefore, not only $\bm{W}^{\prime(:,n_k)}_{i}$ but also $\bm{W}^{\prime(m_k,:)}_{i+1}$ should be defined. Here $m_k$ for the $i$+1-th layer is equal to $n_k$ for the $i$-th layer. The definition for $\bm{W}^{\prime(m_k,:)}_{i+1}$ is shown in Eq. \ref{eq_wo} where $:$ is the wild card for all the output channels.

\begin{equation}
\small
\bm{W}^{\prime(m_k,:)}_{i+1} = \bm{W}^{(m,:)}_{i+1}
\label{eq_wo}
\end{equation}

Based on Eq. \ref{eq_o} and Eq. \ref{eq_wo}, we can have the following relationship which means that the channel splitting for the $i$-th layer will not influence the results of the $i$+1-th layer.

\begin{equation}
\small
\sum_{k=0}^{\mathcal{K}-1}\bm{O}^{\prime(m_k)}_{i} * \bm{W}^{\prime(m_k,:)}_{i+1} = \bm{O}^{(m)}_{i} * \bm{W}^{(m,:)}_{i+1}
\label{eq_oi+1}
\end{equation}

\subsection{Channel Splitting for Non-bottleneck Layers}\label{AA}

Before splitting, each activation output $x$ is quantized as $\lfloor \tfrac{x}{s} \rceil \cdot s$ where $s$ is the original quantization precision.
\begin{equation}
\small
x \rightarrow Q(x;s)=\lfloor \dfrac{x}{s} \rceil \cdot s
\label{eq_x}
\end{equation}

After splitting one channel to $\mathcal{K}$ channels, each activation output $x_k$ is quantized as $\lfloor \tfrac{x_k}{s^{\prime}} \rceil \cdot s^{\prime}$ where $s^{\prime}$ is the quantization precision for the split channels.

\begin{equation}
\small
x_k \rightarrow Q(x_k;s^{\prime})=\lfloor \dfrac{x_k}{s^{\prime}} \rceil \cdot s^{\prime}
\label{eq_ss}
\end{equation}

According to Eq. \ref{eq_o}-\ref{eq_b}, $x$ and $x_k$ have the following relationship so that the dynamic range of split channels can be reduced by $\mathcal{K}$ times compared with the original non-split channel. 
\begin{equation}
\small
x_k=\tfrac{x}{\mathcal{K}}+\delta_k
\label{eq_x1}
\end{equation}

Therefore, in the case of same bit budget, the quantization precision for the split channels can be increased by $\mathcal{K}$ times.
\begin{equation}
\small
s^{\prime}=\dfrac{s}{\mathcal{K}}
\label{eq_s}
\end{equation}

When we set $\delta_k = (\tfrac{1}{2\mathcal{K}}+\tfrac{k}{\mathcal{K}}-\tfrac{1}{2}) \cdot \tfrac{s}{\mathcal{K}}$, the summation of $Q(x_k;s^{\prime})$ for the split $\mathcal{K}$ channels can be derived as

\begin{equation}
\small
\begin{aligned}
&\sum_{k=0}^{\mathcal{K}-1}\lfloor \dfrac{x_k}{s^{\prime}} \rceil \cdot s^{\prime} \\
=&\sum_{k=0}^{\mathcal{K}-1}\lfloor \dfrac{x+\delta_k \cdot \mathcal{K}}{s} \rceil \cdot s^{\prime} \\
=&\sum_{k=0}^{\mathcal{K}-1}\lfloor \dfrac{x}{s} + \dfrac{1}{2\mathcal{K}} + \dfrac{k}{\mathcal{K}} \rfloor \cdot s^{\prime} \\
=& \lfloor \mathcal{K} \cdot(\dfrac{x}{s} + \dfrac{1}{2\mathcal{K}}) \rfloor \cdot s^{\prime} = \lfloor \dfrac{x}{s^{\prime}} \rceil \cdot s^{\prime}
\end{aligned}
\label{eq_4}
\end{equation}
\noindent where the second last equation in Eq. \ref{eq_4} follows \cite{savchev2003mathematical}. Here we can see that by splitting to $\mathcal{K}$ channels, the effect is exactly same as increasing the quantization precision by $\mathcal{K}$ times for the original channel, which proves that the channel splitting can reduce the quantization error.

Though channel splitting can reduce the quantization error, the number of channels is increased so that the network complexity becomes higher. Therefore, we only split the channels whose quantization error will influence more to the coding gain. The detail framework is shown in Section \ref{44}.

\subsection{Channel Splitting for Bottleneck Layer}

Before splitting, each activation output $y$ is rounded to $\lfloor y \rceil$. After splitting one channel to $\mathcal{K}$ channels, each activation output $y_k$ is rounded to $\lfloor y_k \rceil$. When we set $\delta_k$ as $\tfrac{1}{2\mathcal{K}}+\tfrac{k}{\mathcal{K}}-\tfrac{1}{2}$, $\lfloor y \rceil$ and $\lfloor y_k \rceil$ have the following relationship.
\begin{equation}
\small
\begin{aligned}
&\sum_{k=0}^{\mathcal{K}-1}\lfloor y_k \rceil \\
=&\sum_{k=0}^{\mathcal{K}-1}\lfloor \dfrac{y}{\mathcal{K}} + \delta_k \rceil \\
=&\sum_{k=0}^{\mathcal{K}-1}\lfloor \dfrac{y}{\mathcal{K}} + \dfrac{1}{2\mathcal{K}} + \dfrac{k}{\mathcal{K}} \rfloor \\
=&\lfloor \mathcal{K} \cdot(\dfrac{y}{\mathcal{K}} + \dfrac{1}{2\mathcal{K}}) \rfloor = \lfloor y \rceil \\
\end{aligned}
\label{eq_6}
\end{equation}

We can see that the summation of $\lfloor y_k \rceil$ is equal to $\lfloor y \rceil$, which means that the splitting will not influence the coding gain. 

Similar as traditional DCT, there are some low-frequency channels with large magnitude in the bottleneck layer. As a result, if the dynamic range is larger than $2^{b-1}$, the quantization precision will be larger than one according to Eq. \ref{eq_at3}. To ensure that the dynamic range is smaller than $2^{b-1}$, each specific channel can be split to $\mathcal{K}$ channels.
\begin{equation}
\small
\mathcal{K}=\dfrac{max(2^{\lceil log2^{t} \rceil},2^{b-1})}{2^{b-1}}
\label{eq_bottleneck}
\end{equation}

Overall, for the bottleneck layer, the number of incremental channels after the splitting can be calculated as below
\begin{equation}
\small
\sum_{c=0}^{C_b-1}\tfrac{max(2^{\lceil log2^{t_c} \rceil},2^{b-1})-2^{b-1}}{2^{b-1}}
\label{eq_at3_split}
\end{equation}
\noindent where $C_b$ is the number of channels in the bottleneck layer and $t_c$ is the dynamic range for each original channel.

It is noted that after splitting $g^{(3)}_a$, according to Eq. \ref{eq_wo}, we need to modify the weights of $h^{(0)}_a$ to ensure that the outputs of $h^{(0)}_a$ will not be changed. As a result, the hyper decoder outputs the original $\mu, \sigma$ which is not aligned with $\lfloor y_k \rceil$. To address this issue, in the encoding side, we calculate $\sum_{k=0}^{\mathcal{K}-1}\lfloor y_k \rceil$ and then use $\mu, \sigma$ to code the summation result. In the decoding side, we first do the entropy decoding to generate the summation $\sum_{k=0}^{\mathcal{K}-1}\lfloor y_k \rceil$, and then split to $\lfloor y_k \rceil$ as the input of synthesis transform $g_s$.

\subsection{Overall Framework} \label{44}

For a specific layer, supposed that there are $S$ incremental channels after the channel splitting. In order not to increase the overall network complexity, we need to prune the same number of channels. To reduce the coding loss caused by the pruning, we prune the channels with the smallest energy. With larger $S$, more channels will be split so that the quantization error can be reduced. On the other hand, pruning more channels will cause more coding loss. As a result, we need to find an appropriate $S$ to balance the coding gain of splitting and pruning.

\begin{figure}[tb]
	\centering
	\leftskip-0.0cm
	\begin{minipage}[b]{0.4\linewidth}
	\begin{tikzpicture}
	\pgfplotsset{every axis legend/.append style={
			at={(0.2,0.80)},
			anchor=south},every axis y label/.append style={at={(-0.07,0.5)}}}
	
	\begin{axis}[legend columns=4,legend style={font=\tiny},font=\footnotesize,width=5.0cm,height=4cm,xtick=data,xmax=8,legend columns=6,yticklabel style={/pgf/number format/precision=3}]
	\addplot [red, mark=*, mark size=2pt] table {at0.dat};
	\legend{$g^{(0)}_a$}
	\end{axis}
	\end{tikzpicture}
	
	\end{minipage}\hspace{6mm}
	\begin{minipage}[b]{0.4\linewidth}
	\begin{tikzpicture}
	\pgfplotsset{every axis legend/.append style={
			at={(0.2,0.80)},
			anchor=south},every axis y label/.append style={at={(-0.07,0.5)}}}
	\begin{axis}[legend columns=4,legend style={font=\tiny},font=\footnotesize,width=5.0cm,height=4cm,xtick=data,xmax=8,legend columns=6,yticklabel style={/pgf/number format/precision=3}]
	\addplot [blue, mark=*, mark size=2pt] table {at1.dat};
	\legend{$g^{(1)}_a$}
	\end{axis}
	\end{tikzpicture}
	
	\end{minipage}\hspace{6mm}
	\begin{minipage}[b]{0.4\linewidth}
	\begin{tikzpicture}
	\pgfplotsset{every axis legend/.append style={
			at={(0.2,0.80)},
			anchor=south},every axis y label/.append style={at={(-0.07,0.5)}}}
	\begin{axis}[legend columns=4,legend style={font=\tiny},font=\footnotesize,width=5.0cm,height=4cm,xtick=data,xmax=8,legend columns=6,yticklabel style={/pgf/number format/precision=3}]
	\addplot [green, mark=*, mark size=2pt] table {at2.dat};
	\legend{$g^{(2)}_a$}
	\end{axis}
	\end{tikzpicture}
	
	\end{minipage}\hspace{6mm}
	\begin{minipage}[b]{0.4\linewidth}
	\begin{tikzpicture}
	\pgfplotsset{every axis legend/.append style={
			at={(0.2,0.80)},
			anchor=south},every axis y label/.append style={at={(-0.07,0.5)}}}
	\begin{axis}[legend columns=4,legend style={font=\tiny},font=\footnotesize,width=5.0cm,height=4cm,xtick=data,xmax=8,legend columns=6,yticklabel style={/pgf/number format/precision=3}]
	\addplot [cyan, mark=*, mark size=2pt] table {st0.dat};
	\legend{$g^{(0)}_s$}
	\end{axis}
	\end{tikzpicture}
	
	\end{minipage}\hspace{6mm}
	\begin{minipage}[b]{0.4\linewidth}
	\begin{tikzpicture}
	\pgfplotsset{every axis legend/.append style={
			at={(0.2,0.80)},
			anchor=south},every axis y label/.append style={at={(-0.07,0.5)}}}
	\begin{axis}[legend columns=4,legend style={font=\tiny},font=\footnotesize,width=5.0cm,height=4cm,xtick=data,xmax=8,legend columns=6,yticklabel style={/pgf/number format/precision=3}]
	\addplot [Magenta, mark=*, mark size=2pt] table {st1.dat};
	\legend{$g^{(1)}_s$}
	\end{axis}
	\end{tikzpicture}
	
	\end{minipage}\hspace{6mm}
	\begin{minipage}[b]{0.4\linewidth}
	\begin{tikzpicture}
	\pgfplotsset{every axis legend/.append style={
			at={(0.2,0.80)},
			anchor=south},every axis y label/.append style={at={(-0.07,0.5)}}}
	\begin{axis}[legend columns=4,legend style={font=\tiny},font=\footnotesize,width=5.0cm,height=4cm,xtick=data,xmax=8,legend columns=6,yticklabel style={/pgf/number format/precision=3}]
	\addplot [Plum, mark=*, mark size=2pt] table {st2.dat};
	\legend{$g^{(2)}_s$}
	\end{axis}
	\end{tikzpicture}
	
	\end{minipage}\hspace{6mm}
	
	\caption{R-D cost in the case of quantizing each individual layer. Horizontal axis is the number of channels for splitting and pruning, and vertical axis is the corresponding R-D cost. The optimal number of channels for the splitting is 0,2,5,4,4,6 for each layer respectively.}
	\label{fig_split_prune}
\end{figure}
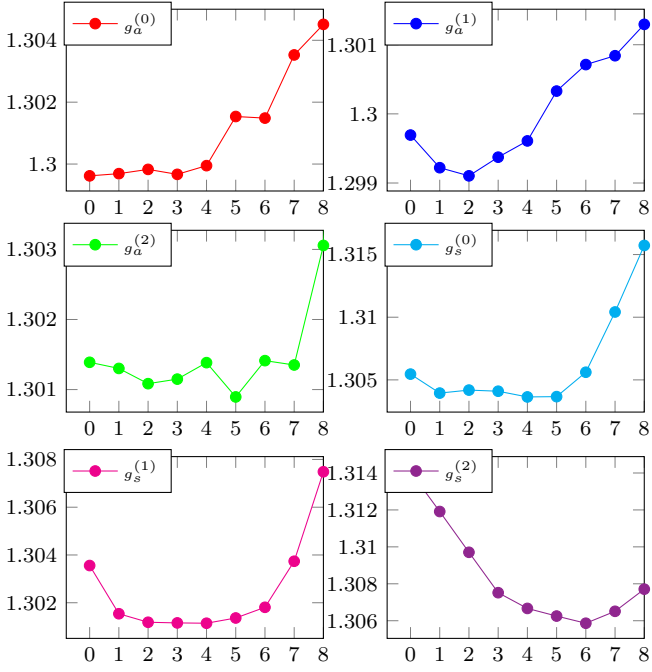

The overall framework is shown in \textit{Algorithm}\ref{alg_framework}. Based on a pre-trained model, for the non-bottleneck layer, the optimal $S$ can be decided by comparing the R-D cost. R-D cost in the case of quantizing each individual layer is shown in Fig. \ref{fig_split_prune}. The horizontal axis is the number of channels which are split to two channels, and the vertical axis is the corresponding R-D cost. We can see that R-D cost for the last five cases can be decreased at first, which means that the coding gain due to the channel splitting is larger than the coding loss caused by the channel pruning. However, when we further split more channels, R-D cost increases since the coding loss from channel pruning will become larger than the coding gain by the channel splitting. For the first case, R-D cost without any channel splitting is the lowest. Therefore, the number of channels for the splitting is set as 0.

For the bottleneck layer, the number of incremental channels after the splitting is shown in Eq. \ref{eq_at3_split}. To remain the overall number of channels, the same number of channels with the smallest energy will be pruned.

Noted that the output of $g^{(3)}_a$ is not only the input of $g^{(0)}_s$, but also the input of $h^{(0)}_a$. Therefore, after splitting the channels of $g^{(3)}_a$, we need to define the weights of $g^{(0)}_s$ as well as $h^{(0)}_a$ as shown in Eq. \ref{eq_wo}.

\begin{algorithm} [t]
	\small
	\caption{Overall Framework}
	\label{alg_framework}
	\begin{algorithmic}[1]
		
		\State \text{Non-bottleneck layer list} $\mathcal{L}_{nb} = [g^{(0,1,2)}_a,g^{(0,1,2)}_s]$
		\State \text{Bottleneck layer list} $\mathcal{L}_{b} = [g^{(3)}_a]$
		
		\State \text{Pre-trained model:} $M$
		
		\State \textbf{Decide split channels $\hat{S}_{nb}$ for non-bottleneck layers}
		
		\For {$l$ in $\mathcal{L}_{nb}$}
		
		\State Calculate channel-wise energy $\mathcal{E}_k$ and $B_k$	
		
		\For{$S$ in range(16)}
		
		\State Split $\mathrm{argmax} B[:S]$ channels
		
		\State Prune $\mathrm{argmin} \mathcal{E}[:S]$ channels
		
		\State Calculate R-D cost $\emph{J}(S)$ 
		
		\EndFor
		
		\State $\hat{S}_{nb} = \underset{S}{\mathrm{argmin}}\emph{J}(S)$
		
		\EndFor
		
		\State \textbf{Decide split channels $\hat{S}_b$ for the bottleneck layer}
		
		\State $\hat{S}_b=\sum_{c=0}^{C-1}\tfrac{max(2^{\lceil log2^t \rceil},2^{b-1})-2^{b-1}}{2^{b-1}}$
		
		\State \textbf{Split and prune the pre-trained model $M$}
		
	\end{algorithmic}
\end{algorithm}

\section{Experimental Results}

\subsection{Training and Network Details}\label{51}

Same as previous works, we train the network with the rate-distortion cost function $\emph{J}= \emph{R}+\lambda \cdot \emph{D}$. The distortion term in the loss is MSE($x,\hat{x}$) when optimizing PSNR, and it is 1-MS-SSIM($x,\hat{x}$) when optimizing the MS-SSIM. Since different $\lambda$ can achieve various trade-off between distortion and rate, we train several models to obtain a R-D curve. When optimizing MSE, $\lambda$ is selected from \{0.001625,0.00325,0.0075,0.015,0.03,0.05\}. When optimizing MS-SSIM, $\lambda$ is picked up from \{3,5,10,40,80,128\}. To get the baseline pre-trained model in \textit{Algorithm 1}, we train about $1 \times 10^6$ iterations, the learning rate is set as $1 \times 10^{-4}$ at first and decayed to $1 \times 10^{-5}$ for the final 80K iterations.

To decide $\hat{S}_{nb}$ and $\hat{S}_{b}$ in \textit{Algorithm 1}, we randomly picked up nine images from ImageNet. For the test set, we use 24 lossless Kodak images \cite{franzen1999kodak} which are commonly used in the previous works. Noted that there is no overlap between the training set and the test set.

\begin{table}[htb]
	\setlength{\abovecaptionskip}{-0.5cm}   
	\setlength{\belowcaptionskip}{0.0cm}
	\caption{BD-rate (\%) compared with the non-quantized floatint-point baseline. Smaller BD-rate means that the coding loss due to the quantization is smaller.}
	\footnotesize
	\begin{center}
		{
			\begin{tabular}{c|c|c|c|c}
				\hline
				
				\textbf{}
				&\textbf{weight}
				&\textbf{activation}
				&\textbf{MSE model}
				&\textbf{MS-SSIM model}	\\ \hline \hline
				
				\text{Ours}
				&8bit
				&8bit
				&4.98
				&4.34 	\\ \hline
				
				\text{\cite{sun2021learned}}
				&8bit
				&8bit
				&5.59
				&9.08 	\\ \hline
				
				\text{\cite{hong2020efficient}}
				&8bit
				&10bit
				&26.5
				&N/A	\\ \hline
				
				\text{\cite{hong2020efficient}}
				&8bit
				&16bit
				&17.9
				&N/A	\\ \hline
				
			\end{tabular}
		}
	\end{center}
	\label{table_all8}
	\vspace{-4mm}
\end{table}

\subsection{Coding Gain Analysis}\label{52}

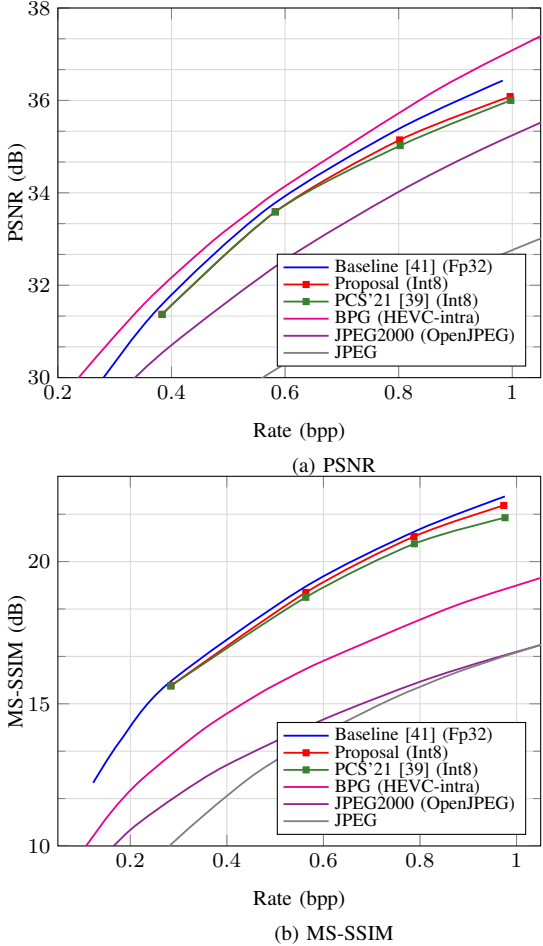
\begin{figure}[H]
	\footnotesize
	\centering
	\begin{minipage}{1.0\linewidth}
		\pgfplotsset{
		}
		\pgfplotsset{every axis plot/.append style={line width=0.7pt}}
		\tikzset{every mark/.append style={scale=0.5}}
		\begin{tikzpicture}
			\begin{axis}[
				width=8cm, height=6.5cm,
				x label style={at={(axis description cs:0.5,-0.1)},anchor=north},
				y label style={at={(axis description cs:-0.05,.5)},anchor=south},
				y tick label style = { /pgf/number format/.cd, precision=3, /tikz/.cd},
				xlabel = {Rate (bpp)},
				ylabel = {PSNR (dB)},
				xmin = 0.20,
				xmax = 1.05,
				ymin = 30.0,
				ymax = 38.0,
				minor y tick num = 2,
				grid = both,
				grid style = {gray!30},
				legend entries = {Baseline\cite{cheng2019deep} (Fp32), Proposal (Int8), PCS'21 \cite{sun2021learned} (Int8), BPG (HEVC-intra), JPEG2000 (OpenJPEG), JPEG},
				legend style={font=\fontsize{7}{7}\selectfont, row sep=-3.5pt, at={(0.5,-0.26)}, anchor=north, inner xsep=2pt,inner ysep=0pt},
				legend cell align=left,
				legend pos = {south east},
				]
				\addplot[smooth, blue, mark size=2.5pt] table {kodak-orig-psnr.dat};
				\addplot[smooth, red, mark=square*] table {kodak-proposal-lq-all-8b-psnr.dat};
				\addplot[smooth, OliveGreen, mark=square*] table {kodak-pcs-lq-all-8b-psnr.dat};
				\addplot[smooth, Magenta] table {kodak-bpg-psnr.dat};
				\addplot[smooth, Plum] table {kodak-jpeg2k-psnr.dat};
				\addplot[smooth, gray] table {kodak-jpeg-psnr.dat};
				
			\end{axis}
		\end{tikzpicture}
		
		\centerline{(a) PSNR}
		
	\end{minipage}\hspace{10mm}
	\begin{minipage}{1.0\linewidth}
		\pgfplotsset{
		}
		\pgfplotsset{every axis plot/.append style={line width=0.7pt}}
		\tikzset{every mark/.append style={scale=0.5}}
		\begin{tikzpicture}
			\begin{axis}[
				width=8cm, height=6.5cm,
				x label style={at={(axis description cs:0.5,-0.1)},anchor=north},
				y label style={at={(axis description cs:-0.05,.5)},anchor=south},
				y tick label style = { /pgf/number format/.cd, precision=3, /tikz/.cd},
				xlabel = {Rate (bpp)},
				ylabel = {MS-SSIM (dB)},
				xmin = 0.05,
				xmax = 1.05,
				ymin = 10.0,
				ymax = 23.0,
				minor y tick num = 2,
				grid = both,
				grid style = {gray!30},
				legend entries = {Baseline\cite{cheng2019deep} (Fp32), Proposal (Int8), PCS'21 \cite{sun2021learned} (Int8), BPG (HEVC-intra), JPEG2000 (OpenJPEG), JPEG},
				legend style={font=\fontsize{7}{7}\selectfont, row sep=-3.5pt, at={(0.5,-0.26)}, anchor=north, inner xsep=2pt,inner ysep=0pt},
				legend cell align=left,
				legend pos = {south east},
				]
				\addplot[smooth, blue, mark size=2.5pt] table {kodak-orig-msssim.dat};
				\addplot[smooth, red, mark=square*] table {kodak-proposal-lq-all-8b-msssim.dat};
				\addplot[smooth, OliveGreen, mark=square*] table {kodak-pcs-lq-all-8b-msssim.dat};
				\addplot[smooth, Magenta] table {kodak-bpg-msssim.dat};
				\addplot[smooth, Plum] table {kodak-jpeg2k-msssim.dat};
				\addplot[smooth, gray] table {kodak-jpeg-msssim.dat};
			\end{axis}
		\end{tikzpicture}
		
		\centerline{(b) MS-SSIM}
		
	\end{minipage}
	\caption{Coding gain comparison with recent work \cite{sun2021learned} for a fully 8-bit QLIC framework.}
	\label{fig_quanall}
	
\end{figure}

We compare the coding gain with two recent QLIC works \cite{sun2021learned,hong2020efficient}. For \cite{sun2021learned}, the author remained the activation in the main path unquantized. When quantizing the activation in the main path to 8-bit, the results are shown in Fig. \ref{fig_quanall}. Compared with \cite{sun2021learned}, the BD-rate \cite{bjontegaard2001calculation} compared with the floating-point anchor can be reduced by 0.61\% (5.59-4.98) and 4.74\% (9.08-4.34) for MSE model and MS-SSIM model respectively. We can achieve more coding gain for MS-SSIM model since there are more channels for the splitting and pruning by following \textit{Algorithm 1}.

\cite{hong2020efficient} also gave the BD-rate compared with the floating-point anchor for the MSE-optimized model. In the case of 8-bit weight quantization and 10-bit activation quantization, BD-rate is 26.5\%. In the case of 8-bit weight quantization and 16-bit activation quantization, BD-rate is 17.9\%. Compared with \cite{hong2020efficient}, our coding loss caused by the quantization is much smaller.

\cite{balle2018integer} also illustrate that there will be a coding loss after the quantization. However, the bit-width of weight and activation is not given in the paper. For the fairness, we did not compare with \cite{balle2018integer} here.

\begin{figure*}[h]
	\footnotesize
	\centering
	\leftskip-0.0cm
	\begin{minipage}[b]{0.2\linewidth}
		\pgfplotsset{
		}
		\pgfplotsset{every axis plot/.append style={line width=0.7pt}}
		\tikzset{every mark/.append style={scale=0.5}}
		\begin{tikzpicture}
			\begin{axis}[
				width=5.5cm, height=5cm,
				x label style={at={(axis description cs:0.5,-0.1)},anchor=north},
				y label style={at={(axis description cs:-0.05,.5)},anchor=south},
				y tick label style = { /pgf/number format/.cd, precision=3, /tikz/.cd},
				xlabel = {Rate (bpp)},
				ylabel = {PSNR (dB)},
				xmin = 0.2,
				xmax = 1.05,
				ymin = 30.0,
				ymax = 37.0,
				minor y tick num = 2,
				grid = both,
				grid style = {gray!30},
				legend entries = {Fp32 baseline, w/ proposal, w/o proposal},
				legend style={font=\fontsize{7}{7}\selectfont, row sep=-3.5pt, at={(0.5,-0.26)}, anchor=north, inner xsep=2pt,inner ysep=0pt},
				legend cell align=left,
				legend pos = {south east},
				]
				\addplot[smooth, blue, mark size=2.5pt] table {kodak-orig-psnr.dat};
				\addplot[smooth, OliveGreen, mark=square*] table {kodak-proposal-lq-main-8b-psnr.dat};
				\addplot[smooth, red, mark=square*] table {kodak-orig-lq-main-8b-psnr.dat};
				
			\end{axis}
		\end{tikzpicture}
		
		\centerline{(a) Kodak, PSNR, 8-bit}
		
	\end{minipage}\hspace{8mm}
	\vspace{5mm}
	\begin{minipage}[b]{0.2\linewidth}
		\pgfplotsset{
		}
		\pgfplotsset{every axis plot/.append style={line width=0.7pt}}
		\tikzset{every mark/.append style={scale=0.5}}
		\begin{tikzpicture}
			\begin{axis}[
				width=5.5cm, height=5cm,
				x label style={at={(axis description cs:0.5,-0.1)},anchor=north},
				y label style={at={(axis description cs:-0.05,.5)},anchor=south},
				y tick label style = { /pgf/number format/.cd, precision=3, /tikz/.cd},
				xlabel = {Rate (bpp)},
				ylabel = {PSNR (dB)},
				xmin = 0.2,
				xmax = 1.05,
				ymin = 28.0,
				ymax = 37.0,
				minor y tick num = 2,
				grid = both,
				grid style = {gray!30},
				legend entries = {Fp32 baseline, w/ proposal, w/o proposal},
				legend style={font=\fontsize{7}{7}\selectfont, row sep=-3.5pt, at={(0.5,-0.26)}, anchor=north, inner xsep=2pt,inner ysep=0pt},
				legend cell align=left,
				legend pos = {south east},
				]
				
				\addplot[smooth, blue, mark size=2.5pt] table {kodak-orig-psnr.dat};
				\addplot[smooth, OliveGreen, mark=triangle*] table {kodak-proposal-lq-main-7b-psnr.dat};		
				\addplot[smooth, red, mark=triangle*] table {kodak-orig-lq-main-7b-psnr.dat};
				
			\end{axis}
		\end{tikzpicture}
		
		\centerline{(b) Kodak, PSNR, 7-bit}
		
	\end{minipage}\hspace{8mm}
	\begin{minipage}[b]{0.2\linewidth}
		\pgfplotsset{
		}
		\pgfplotsset{every axis plot/.append style={line width=0.7pt}}
		\tikzset{every mark/.append style={scale=0.5}}
		\begin{tikzpicture}
			\begin{axis}[
				width=5.5cm, height=5cm,
				x label style={at={(axis description cs:0.5,-0.1)},anchor=north},
				y label style={at={(axis description cs:-0.05,.5)},anchor=south},
				y tick label style = { /pgf/number format/.cd, precision=3, /tikz/.cd},
				xlabel = {Rate (bpp)},
				ylabel = {MS-SSIM (dB)},
				xmin = 0.10,
				xmax = 1.05,
				ymin = 11.0,
				ymax = 23.0,
				minor y tick num = 2,
				grid = both,
				grid style = {gray!30},
				legend entries = {Fp32 baseline, w/ proposal, w/o proposal},
				legend style={font=\fontsize{7}{7}\selectfont, row sep=-3.5pt, at={(0.5,-0.26)}, anchor=north, inner xsep=2pt,inner ysep=0pt},
				legend cell align=left,
				legend pos = {south east},
				]		
				\addplot[smooth, blue, mark size=2.5pt] table {kodak-orig-msssim.dat};
				\addplot[smooth, OliveGreen, mark=square*] table {kodak-proposal-lq-main-8b-msssim.dat};
				\addplot[smooth, red, mark=square*] table {kodak-orig-lq-main-8b-msssim.dat};
			\end{axis}
		\end{tikzpicture}
		
		\centerline{(c) Kodak, MS-SSIM, 8-bit}
		
	\end{minipage}\hspace{8mm}
	\begin{minipage}[b]{0.2\linewidth}
		\pgfplotsset{
		}
		\pgfplotsset{every axis plot/.append style={line width=0.7pt}}
		\tikzset{every mark/.append style={scale=0.5}}
		\begin{tikzpicture}
			\begin{axis}[
				width=5.5cm, height=5cm,
				x label style={at={(axis description cs:0.5,-0.1)},anchor=north},
				y label style={at={(axis description cs:-0.05,.5)},anchor=south},
				y tick label style = { /pgf/number format/.cd, precision=3, /tikz/.cd},
				xlabel = {Rate (bpp)},
				ylabel = {MS-SSIM (dB)},
				xmin = 0.10,
				xmax = 1.05,
				ymin = 11.0,
				ymax = 23.0,
				minor y tick num = 2,
				grid = both,
				grid style = {gray!30},
				legend entries = {Fp32 baseline, w/ proposal, w/o proposal},
				legend style={font=\fontsize{7}{7}\selectfont, row sep=-3.5pt, at={(0.5,-0.26)}, anchor=north, inner xsep=2pt,inner ysep=0pt},
				legend cell align=left,
				legend pos = {south east},
				]
				
				\addplot[smooth, blue, mark size=2.5pt] table {kodak-orig-msssim.dat};
				\addplot[smooth, OliveGreen, mark=triangle*] table {kodak-proposal-lq-main-7b-msssim.dat};		
				\addplot[smooth, red, mark=triangle*] table {kodak-orig-lq-main-7b-msssim.dat};
				
			\end{axis}
		\end{tikzpicture}
		
		\centerline{(d) Kodak, MS-SSIM, 7-bit}
		
	\end{minipage}
	
	\begin{minipage}[b]{0.2\linewidth}
		\pgfplotsset{
		}
		\pgfplotsset{every axis plot/.append style={line width=0.7pt}}
		\tikzset{every mark/.append style={scale=0.5}}
		\begin{tikzpicture}
			\begin{axis}[
				width=5.5cm, height=5cm,
				x label style={at={(axis description cs:0.5,-0.1)},anchor=north},
				y label style={at={(axis description cs:-0.05,.5)},anchor=south},
				y tick label style = { /pgf/number format/.cd, precision=3, /tikz/.cd},
				xlabel = {Rate (bpp)},
				ylabel = {PSNR (dB)},
				xmin = 0.2,
				xmax = 0.95,
				ymin = 30.0,
				ymax = 38.0,
				minor y tick num = 2,
				grid = both,
				grid style = {gray!30},
				legend entries = {Fp32 baseline, w/ proposal, w/o proposal},
				legend style={font=\fontsize{7}{7}\selectfont, row sep=-3.5pt, at={(0.5,-0.26)}, anchor=north, inner xsep=2pt,inner ysep=0pt},
				legend cell align=left,
				legend pos = {south east},
				]
				\addplot[smooth, blue, mark size=2.5pt] table {CLIC-orig-psnr.dat};
				\addplot[smooth, OliveGreen, mark=triangle*] table {clic-proposal-lq-main-8b-psnr.dat};		
				\addplot[smooth, red, mark=triangle*] table {clic-orig-lq-main-8b-psnr.dat};
			\end{axis}
		\end{tikzpicture}
		
		\centerline{(e) CLIC, PSNR, 8-bit}
		
	\end{minipage}\hspace{8mm}
	\begin{minipage}[b]{0.2\linewidth}
		\pgfplotsset{
		}
		\pgfplotsset{every axis plot/.append style={line width=0.7pt}}
		\tikzset{every mark/.append style={scale=0.5}}
		\begin{tikzpicture}
			\begin{axis}[
				width=5.5cm, height=5cm,
				x label style={at={(axis description cs:0.5,-0.1)},anchor=north},
				y label style={at={(axis description cs:-0.05,.5)},anchor=south},
				y tick label style = { /pgf/number format/.cd, precision=3, /tikz/.cd},
				xlabel = {Rate (bpp)},
				ylabel = {PSNR (dB)},
				xmin = 0.1,
				xmax = 0.95,
				ymin = 30.0,
				ymax = 38.0,
				minor y tick num = 2,
				grid = both,
				grid style = {gray!30},
				legend entries = {Fp32 baseline, w/ proposal, w/o proposal},
				legend style={font=\fontsize{7}{7}\selectfont, row sep=-3.5pt, at={(0.5,-0.26)}, anchor=north, inner xsep=2pt,inner ysep=0pt},
				legend cell align=left,
				legend pos = {south east},
				]
				\addplot[smooth, blue, mark size=2.5pt] table {CLIC-orig-psnr.dat};
				\addplot[smooth, OliveGreen, mark=triangle*] table {clic-proposal-lq-main-7b-psnr.dat};		
				\addplot[smooth, red, mark=triangle*] table {clic-orig-lq-main-7b-psnr.dat};
				
			\end{axis}
		\end{tikzpicture}
		
		\centerline{(f) CLIC, PSNR, 7-bit}
		
	\end{minipage}\hspace{8mm}
	\begin{minipage}[b]{0.2\linewidth}
		\pgfplotsset{
		}
		\pgfplotsset{every axis plot/.append style={line width=0.7pt}}
		\tikzset{every mark/.append style={scale=0.5}}
		\begin{tikzpicture}
			\begin{axis}[
				width=5.5cm, height=5cm,
				x label style={at={(axis description cs:0.5,-0.1)},anchor=north},
				y label style={at={(axis description cs:-0.05,.5)},anchor=south},
				y tick label style = { /pgf/number format/.cd, precision=3, /tikz/.cd},
				xlabel = {Rate (bpp)},
				ylabel = {MS-SSIM (dB)},
				xmin = 0.1,
				xmax = 1.05,
				ymin = 13.0,
				ymax = 25.0,
				minor y tick num = 2,
				grid = both,
				grid style = {gray!30},
				legend entries = {Fp32 baseline, w/ proposal, w/o proposal},
				legend style={font=\fontsize{7}{7}\selectfont, row sep=-3.5pt, at={(0.5,-0.26)}, anchor=north, inner xsep=2pt,inner ysep=0pt},
				legend cell align=left,
				legend pos = {south east},
				]
				\addplot[smooth, blue, mark size=2.5pt] table {CLIC-orig-msssim.dat};
				\addplot[smooth, OliveGreen, mark=triangle*] table {clic-proposal-lq-main-8b-msssim.dat};		
				\addplot[smooth, red, mark=triangle*] table {clic-orig-lq-main-8b-msssim.dat};
			\end{axis}
		\end{tikzpicture}
		
		\centerline{(g) CLIC, MS-SSIM, 8-bit}
		
	\end{minipage}\hspace{8mm}
	\begin{minipage}[b]{0.2\linewidth}
		\pgfplotsset{
		}
		\pgfplotsset{every axis plot/.append style={line width=0.7pt}}
		\tikzset{every mark/.append style={scale=0.5}}
		\begin{tikzpicture}
			\begin{axis}[
				width=5.5cm, height=5cm,
				x label style={at={(axis description cs:0.5,-0.1)},anchor=north},
				y label style={at={(axis description cs:-0.05,.5)},anchor=south},
				y tick label style = { /pgf/number format/.cd, precision=3, /tikz/.cd},
				xlabel = {Rate (bpp)},
				ylabel = {MS-SSIM (dB)},
				xmin = 0.05,
				xmax = 0.85,
				ymin = 11.0,
				ymax = 25.0,
				minor y tick num = 2,
				grid = both,
				grid style = {gray!30},
				legend entries = {Fp32 baseline, w/ proposal, w/o proposal},
				legend style={font=\fontsize{7}{7}\selectfont, row sep=-3.5pt, at={(0.5,-0.26)}, anchor=north, inner xsep=2pt,inner ysep=0pt},
				legend cell align=left,
				legend pos = {south east},
				]
				\addplot[smooth, blue, mark size=2.5pt] table {CLIC-orig-msssim.dat};
				\addplot[smooth, OliveGreen, mark=triangle*] table {clic-proposal-lq-main-7b-msssim.dat};		
				\addplot[smooth, red, mark=triangle*] table {clic-orig-lq-main-7b-msssim.dat};
			\end{axis}
		\end{tikzpicture}
		
		\centerline{(h) CLIC, MS-SSIM, 7-bit}
		
	\end{minipage}
	\caption{Coding gain comparison of 7/8-bit activation quantization with 32-bit floating-point baseline.}
	\label{fig_codinggain}
	
\end{figure*}
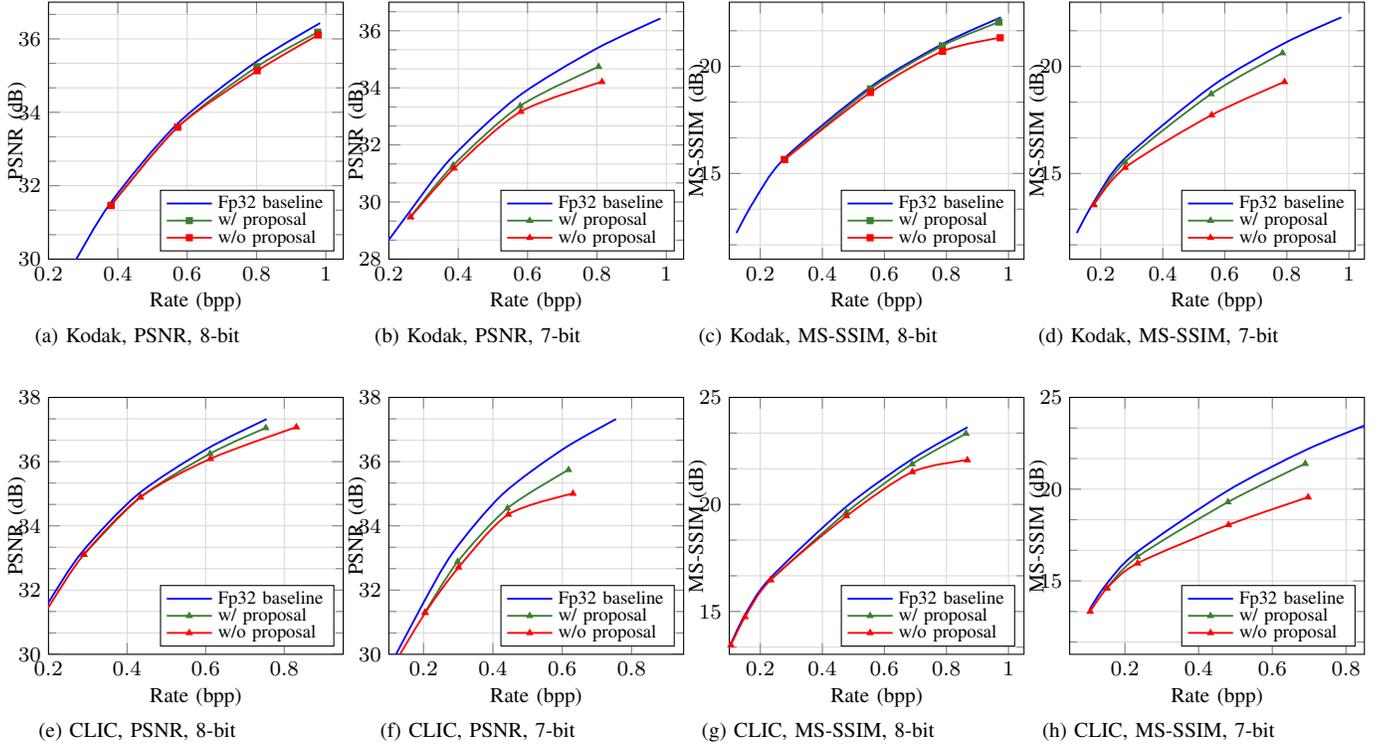

In addition to the comparison with \cite{sun2021learned,hong2020efficient}, we also give an ablation study when using our proposal for the activation quantization in the main path. The results are shown in Fig. \ref{fig_codinggain}. We explore the case of 7-bit and 8-bit quantizations. For the 8-bit quantization, we evaluate four largest bitrate models For the 7-bit quantization, we evaluate four medium bitrate models. Two datasets are used for the evaluation. We can see that the coding gain improvement for the 7-bit quantization is larger than that for the 8-bit quantization. Besides, we can enhance more coding gain for the MS-SSIM models than the MSE models, and this phenomenon is similar as Fig. \ref{fig_quanall}.

\subsection{Effect of Quantization Method}\label{53}

By using the proposal, we can reduce the coding loss of activation quantization without network fine tuning. In fact, our method can be combined with any fine tuning methods to further reduce the coding loss. We evaluate two largest bitrate models of MSE models and MS-SSIM models respectively. Two different bit-width are used in the experiment. From the results in Table \ref{table_ewgs}, we can see that \cite{lee2021network} can reach better coding gain than our proposal when quantizing MSE models. However, when combining our method with \cite{lee2021network}, we can achieve lower R-D cost than \cite{lee2021network}. For the MS-SSIM models, we can achieve much better coding gain than \cite{lee2021network}. In addition, the combination of proposal and \cite{lee2021network} can also lead to a further improvement.

\begin{table}[t]
	\setlength{\abovecaptionskip}{-0.5cm}   
	\setlength{\belowcaptionskip}{0.0cm}
	\caption{R-D cost comparison of our proposal, state-of-the-art network quantization method \cite{lee2021network} and their combination}
	\footnotesize
	\begin{center}
		{
			\begin{tabular}{c|c|c|c|c}
				\hline
				
				\textbf{Model}
				&\textbf{Quan. Bit}
				&\textbf{Method}
				&\textbf{Kodak}
				&\textbf{CLIC}	\\ \hline \hline
				
				\multirow{5}*{0.03}
				&\multirow{5}*{7bit}
				&\text{Baseline}
				&1.4071
				&1.1193 	\\ \cline{3-5}
				
				\textbf{}
				&\text{ }
				&\text{w/o Proposal}
				&1.5975
				&1.3059 	\\ \cline{3-5}
				
				\textbf{}
				&\text{ }
				&\text{Proposal}
				&1.5125
				&1.2119	\\ \cline{3-5}
				
				\textbf{}
				&\text{ }
				&\text{EWGS\cite{lee2021network}}
				&1.4759
				&1.1826 	\\ \cline{3-5}
				
				\textbf{}
				&\text{ }
				&\text{Proposal+EWGS\cite{lee2021network}}
				&1.4478
				&1.1498 	\\ \hline \hline
				
				\multirow{5}*{0.05}
				&\multirow{5}*{8bit}
				&\text{Baseline}
				&1.7818
				&1.4514 	\\ \cline{3-5}
				
				\textbf{}
				&\text{ }
				&\text{w/o Proposal}
				&1.8405
				&1.5710 	\\ \cline{3-5}
				
				\textbf{}
				&\text{ }
				&\text{Proposal}
				&1.8279
				&1.4927	\\ \cline{3-5}
				
				\textbf{}
				&\text{ }
				&\text{EWGS\cite{lee2021network}}
				&1.8178
				&1.4796 	\\ \cline{3-5}
				
				\textbf{}
				&\text{ }
				&\text{Proposal+EWGS\cite{lee2021network}}
				&1.8106
				&1.4752 	\\ \hline \hline
				
				\multirow{5}*{80}
				&\multirow{5}*{7bit}
				&\text{Baseline}
				&1.4064
				&1.1782 	\\ \cline{3-5}
				
				\textbf{}
				&\text{ }
				&\text{w/o Proposal}
				&1.7370
				&1.5827 	\\ \cline{3-5}
				
				\textbf{}
				&\text{ }
				&\text{Proposal}
				&1.4788
				&1.2706	\\ \cline{3-5}
				
				\textbf{}
				&\text{ }
				&\text{EWGS\cite{lee2021network}}
				&1.5418
				&1.3484		\\ \cline{3-5}
				
				\textbf{}
				&\text{ }
				&\text{Proposal+EWGS\cite{lee2021network}}
				&1.4341
				&1.2120 	\\ \hline \hline
				
				\multirow{5}*{128}
				&\multirow{5}*{8bit}
				&\text{Baseline}
				&1.7267
				&1.4203 	\\ \cline{3-5}
				
				\textbf{}
				&\text{ }
				&\text{w/o Proposal}
				&1.9138
				&1.6617 	\\ \cline{3-5}
				
				\textbf{}
				&\text{ }
				&\text{Proposal}
				&1.7640
				&1.4608	\\ \cline{3-5}
				
				\textbf{}
				&\text{ }
				&\text{EWGS\cite{lee2021network}}
				&1.8164
				&1.5348		\\ \cline{3-5}
				
				\textbf{}
				&\text{ }
				&\text{Proposal+EWGS\cite{lee2021network}}
				&1.7469
				&1.4473 	\\ \hline \hline
				
			\end{tabular}
		}
	\end{center}
	\label{table_ewgs}
	\vspace{-4mm}
\end{table}

\section{Conclusion}

This paper gives an activation quantization method for LIC. Not only the activation output of the bottleneck layer, but also non-bottleneck layer are quantized. By channel splitting, the dynamic range of channels can be reduced so that the quantization error can be reduced. However, channel splitting will lead to some incremental channels. To keep the network size as origin, channels with the smallest energy will be pruned. As a result, without any network overhead, the coding loss can be reduced for the quantized LIC. In the future work, we will map the quantized LIC network on some specific hardwares such as FPGA. We will also quantize other LIC network structures.


\ifCLASSOPTIONcaptionsoff
  \newpage
\fi



%

\bibliographystyle{IEEEtran}
\bibliography{strings,refs}

%
%

%








\end{document}